\documentclass[manuscript]{aastex}

\usepackage{url}
\shorttitle{solarflare.njit.edu}
\shortauthors{Sadykov et al.}

\begin{document}

\title{INTERACTIVE MULTI-INSTRUMENT DATABASE OF SOLAR FLARES}
\author{Viacheslav M Sadykov\altaffilmark{1,2}, Alexander G Kosovichev\altaffilmark{1,2,3}, Vincent Oria\altaffilmark{1,4}, Gelu M Nita\altaffilmark{1,2,5}}

\affil{$^1$Center for Computational Heliophysics, New Jersey Institute of Technology, Newark, NJ 07102, USA}
\affil{$^2$Department of Physics, New Jersey Institute of Technology, Newark, NJ 07102, USA}
\affil{$^3$NASA Ames Research Center, Moffett Field, CA 94035, USA}
\affil{$^4$Department of Computer Sciences, New Jersey Institute of Technology, Newark, NJ 07102, USA}
\affil{$^5$Center for Solar-Terrestrial Research, New Jersey Institute of Technology, Newark, NJ 07102, USA}

\begin{abstract}
Solar flares are complicated physical phenomena that are observable in a broad range of the electromagnetic spectrum, from radiowaves to $\gamma$-rays. For a more comprehensive understanding of flares, it is necessary to perform a combined multi-wavelength analysis using observations from many satellites and ground-based observatories. For efficient data search, integration of different flare lists and representation of observational data, we have developed an Interactive Multi-Instrument Database of Solar Flares (\url{https://solarflare.njit.edu/}). The web accessible database is fully functional and allows the user to search for uniquely-identified flare events based on their physical descriptors and availability of observations by a particular set of instruments. Currently, the data from three primary flare lists (GOES, RHESSI and HEK) and a variety of other event catalogs (Hinode, Fermi GBM, Konus-Wind, OVSA flare catalogs, CACTus CME catalog, Filament eruption catalog) and observing logs (IRIS and Nobeyama coverage) are integrated, and an additional set of physical descriptors (temperature and emission measure) is provided along with an observing summary, data links, and multi-wavelength light curves for each flare event since January, 2002. We envision that this new tool will allow researchers to significantly speed up the search of events of interest for statistical and case studies.
\end{abstract}


\section{Introduction}
\label{section_intro}

Among many interesting heliospheric phenomena solar flares and related events (Coronal Mass Ejections(CMEs), Solar Energetic Particles (SEP), Hard X-ray and gamma radiation) are of particular interest because they can cause many problems for the terrestrial environment, such as initiating return currents and breaking power grids at high latitudes, disturbing the magnetosphere, and damaging satellites equipment. For better understanding of solar flares and their prediction, it is crucially important to analyze multi-wavelength observations, as different physical processes are reflected at different energies. For example, in the Standard flare model, hard X-ray radiation represents bremsstrahlung emission of accelerated particles, and carries information about acceleration processes associated with magnetic energy release. At the same time, observations of visible and ultraviolet spectral lines allow us to understand photospheric and chromospheric responses to the energy release.

Flare events are observed by a variety of space- and ground-based instruments in different wavelengths. Usually, flare lists are created for specific routinely-observing instruments. Currently, the primary flare catalog is based on soft X-ray emission peaks (so-called X-ray flare classes) observed by the Geostationary Operational Environmental Satellites~\citep[GOES,][]{Bornmann96}. The GOES X-ray instruments have observed the solar activity for several decades, and created the largest database of solar flares. Another example of flare-observing instruments is the Reuven Ramaty High Energy Solar Spectroscopic Imager~\citep[RHESSI,][]{Lin02}, launched in 2002. RHESSI observes the X-ray radiation of flares in a wide range of energies, from 6\,keV to $>$\,300\,keV. The satellite detects the events and has its own flare list separate from the GOES flare list. Two instruments onboard the Solar Dynamics Observatory (SDO) observe solar flares in EUV bands: the Extreme Ultraviolet Variability Experiment instrument~\citep[EVE,][]{Woods12} observes the EUV spectra of the integrated solar emission, and the Atmospheric Imaging Assembly~\citep[AIA,][]{Lemen12} instrument observes high-resolution images in several EUV bands. The flare data from both of these instruments are stored in independent data catalogs. In the EVE data, the flares are detected as enhancement of the EUV emission, and in the AIA data, the flare events are detected using image processing algorithms~\citep{Martens12} and are summarized in the Heliophysics Event Knowledgebase~\citep{Hurlburt12}. Some of these events are linked to the GOES database.

The Virtual Solar Observatory (VSO, \url{http://sdac.virtualsolar.org/cgi/search}) collects and accesses metadata from many space missions and ground-based observatories, allowing the user to search for available data for a particular time range. It also contains several flare and flare-related event catalogs (SOHO/LASCO CME catalog, GOES X-ray Catalog, RHESSI Flare list etc.) The user can search the events having particular properties within each catalog, and request the corresponding data. However, VSO does not allow a detailed search based on flare parameters. The RHESSI browser (\url{http://sprg.ssl.berkeley.edu/~tohban/browser/}) allows users to look at RHESSI and Fermi data products, and to check the observational coverage of the detected flare events by the Hinode and IRIS satellites. However, this browser does not provide the ability to search for flares having particular properties.

Many problems of flare physics require performing analyses using data from a particular set of instruments, or/and for a sample of flares with particular characteristics. For example, find all events having GOES class $\geq$ C5.0 that were observed by RHESSI, or find the flares observed by the IRIS satellite in the EUV range, and by the Nobeyama radio telescope, in the microwave range. To address this type of problems, we developed a new interactive multi-instrument database of solar flares.

This is not the first attempt to provide flare lists or event catalogs. For example, the Owens Valley Solar Array \citep[OVSA,][]{Hurford1984,Gary1990} legacy radio bursts database \citep{Nita2004} allows for the searching of events based on their physical parameter ranges, and the Solar Flare Finder tool \citep[][\url{http://hesperia.gsfc.nasa.gov/sff/}]{Milligan17}, recently developed as a part of Solar SoftWare package for Interactive Data Language \citep[SSW IDL,][]{Freeland00}, allows selecting flaring events simultaneously observed by GOES, RHESSI, SDO/AIA, Hinode, SDO/EVE and IRIS, and to see their data summaries. However there is still room for a comprehensive solution that could include features such as implementing user-interactive filters, providing more convenient representation of the output set of events and improving the accessibility of the data by providing a search tool with minimum software requirements. We provide these and other new features in our newly developed database.

\begin{figure}[p]
	\centering
	\includegraphics[width=0.8\linewidth]{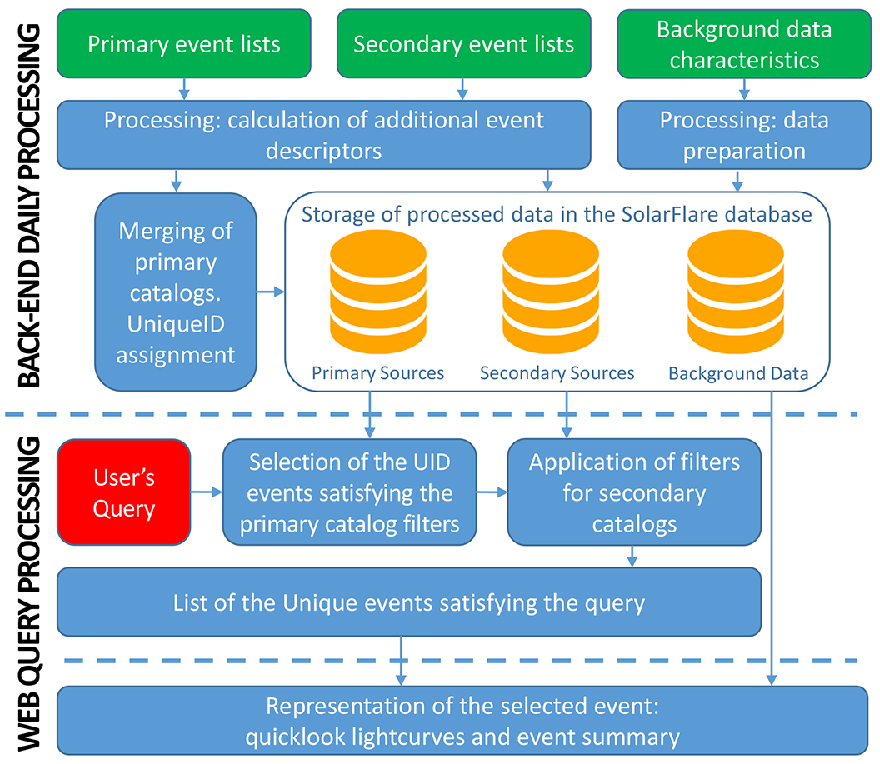}
	\caption{Schematic representation of the Interactive Multi-Instrument Database of Solar Flares (IMIDSF). The database stores the metadata from GOES, RHESSI, SDO, SOHO, Hinode, IRIS, Fermi and other space- and ground-based instruments, as well as some instrument-specific light curves.}
	\label{schema}
\end{figure}

Figure~\ref{schema} represents the basic structure of the database, and each block of this Figure is explained in the following sections of this paper. In Sec.~\ref{section_data}, we describe the flare and flare-related event lists, as well as actual data, which serve as daily input for our database. In Sec.~\ref{section_backend}, we explain the daily processing of the event lists and data: integration of the flares from different lists, calculation of additional event descriptors, preparation (smoothing) of the light curves. The processed data are stored in a MySQL database allowing convenient and fast interaction. In Sec.~\ref{section_query}, we describe the web interface, the structure and logic of queries for our database, and the structure of the output data available to the user. A query example is also presented in this section. We present our conclusions in Sec.~\ref{section_conclusion}.

\section{Data collection and Storage}
\label{section_data}

In this Section, we describe the catalogs of events used as inputs. The complete and up-to-date list of the integrated event sources can be found via the link \url{https://solarflare.njit.edu/datasources.html}, and the current list is summarized in Table~\ref{table_datalinks}.

\subsection{Primary Event Lists}

The event lists are divided in ``primary'' and ``secondary''. The primary event lists represent daily-updated lists of flares independently detected by GOES, RHESSI and SDO/AIA instruments. The secondary lists include partial flare lists, representing subsets of the primary lists, and catalogs of flare-related phenomena (such as Filament eruptions or CMEs).

Each event in the primary lists is characterized by start, peak, and end times. In most cases, the event coordinates and the associated active region number are also reported. We use three primary event catalogs:

\begin{itemize}
	\item \textit{GOES flare list}. The daily lists of events observed by the GOES satellites in the 1-8\,{\AA} channel are available from June 2002 to present. The reported characteristics include the GOES class, the X-ray peak flux during the event, and the information about the active region and coordinates of the event (not for all events). The daily lists are available at the NOAA website.
	\item \textit{RHESSI flare list.} The list of the flares observed by the RHESSI X-ray telescope from February 2002 to present. Besides the usual descriptors (the flare times and position), the catalog contains the highest energy band in which the flares were observed, the number of counts during the flares, and a variety of observational quality flags.
	\item \textit{HEK SDO/AIA event list}. The events detected in the EUV images from the AIA/SDO instrument from February 2010 to present. The events reported in this catalog are characterized by a variety of different parameters (besides the common ones): the wavelength in which the event was detected, the coordinates in a variety of coordinate systems, peak fluxes, web links to quick-look images and movies, etc.
\end{itemize}

These three primary event lists are integrated into a single database, and Unique Identifiers (UniqueID) are prescribed for each event, as discussed in Sec.~\ref{section_backend}.

\subsection{Secondary event sources.}

In addition to the primary event lists, the following secondary data sources are integrated in our catalog:

\begin{itemize}
	\item \textit{The Interface Region Imaging Spectrograph data}~\citep[IRIS,][]{DePontieu14}. IRIS obtains the slit-jaw UV images, as well as spectra of the Sun. The flare events are associated with IRIS observations based on the time and pointing stored in the form of instrument observing logs. The quicklook data web links allow the users to select the events of interest.
	\item \textit{Hinode flare catalog}~\citep{Watanabe12}. The original catalog includes the events from the GOES flare list observed by the Hinode spacecraft. This catalog includes the availability of observations for each Hinode instrument, and quicklook data links.
	\item \textit{Fermi Gamma-ray Burst Monitor~\citep[GBM,][]{Meegan09} solar flare catalog}. The list of the flares observed by the Fermi GBM in the 8\,keV-40\,MeV energy range from November 2008 to present. This catalog includes duration of the observed flares and number of counts during the flares.
	\item \textit{Nobeyama Radio-polarimeter light curves~\citep{Nakajima94}.} The polarimetric measurements from Nobeyama Radio Observatory are available for almost every day, usually approximately 8 hours per day.
	\item \textit{OVSA  legacy flare catalog} \citep{Nita2004}, which includes short-time summaries of events observed by the Ovens Valley Solar Array in the 1-18 GHz microwave range, from 2001 to 2003 only.
	\item \textit{Computer Aided CME Tracking (CACTus) catalog}~\citep{Robbrecht04,Robbrecht09}. This catalog collects records of CMEs detected by the LASCO/SOHO coronograph, and contains a variety of CME properties, including the onset time, principal angle, velocity, etc. The CME-flare event matching algorithm currently implemented in our database is based on the following rule: the recorded CME onset time must lie in an predefined time interval relative to the flare start and end times, which can be interactively adjusted by the user by means of a ``Search time interval'' filter.
	\item \textit{Filament eruption catalog}~\citep{McCauley15}. The filament-flare event matching is based on the time and position of the eruptions. A variety of filament parameters are available. The catalog production was stopped on Oct, 19, 2014.
	\item \textit{Konus-Wind flare catalog}~\citep{Aptekar95,Palshin14}. The original catalog includes the events from the GOES flare list observed by the Konus-Wind spacecraft.
\end{itemize}

Our database is designed in such a way that it can support a continuously expanding number of input sources of different types. These include flare and flare-related event catalogs, and information about the observational coverage by different instruments.

\subsection{Background Data Characteristics}

The aim of the developed database is not only to collect and integrate the flare records from different sources, but also to provide users with an overview of the events they potentially want to study. The flare catalogs themselves already contain many useful quicklook links. For example, each HEK flare record contains the links to the quicklook movies and images obtained by AIA/SDO. Our approach is to contribute to the flare overview, and present additional data for each selected event.

Here is the summary of the time plots we provide for each event (if covered by the instrument):

\begin{itemize}
	\item \textit{GOES X-ray light curves} (two channels 0.5-4\,{\AA} and 1-8\,{\AA}).
	\item \textit{Temperature and Emission Measure determined from the GOES X-ray data in a one-temperature approximation.}
	\item \textit{SDO/EVE ESP light curves} (four diode channels: 18\,nm, 26\,nm, 30\,nm, 36\,nm).
	\item \textit{Nobeyama Polarimeter data} (six frequency bands, two polarizations: 1~GHz, 2~GHz, 3.75~GHz, 9.4~GHz, 17~GHz, 35~GHz, I and V polarizations for each frequency).
\end{itemize}

Each of the described sources is updated daily. The Temperature (T) and Emission Measure (EM) for the events are computed using the Temperature and Emission measure-Based Background Subtraction Algorithm~\citep[TEBBS,][]{Ryan12} described in Section~\ref{section_backend}. For the SDO/EVE ESP light curves, we apply 10-second averaging in order to obtain smoother profiles. The same approach is used for the Nobeyama Polarimeter data. These characteristics which help describe the flare evolution, may provide useful information for selecting particular events for further detailed studies.

\subsection{Data Storage and Queries}

For quick access to the flare catalog information and other metadata derived from the observational data, we store the information in a MySQL database. Each catalog is created as a separate relation, and proper indexes are created to speed up the search. A web interface allows the user to query and visualize the results. A query can take from several seconds (for a typical one-month time period) to several minutes (for the entire time period and no active filters).

\section{Data Enrichment and Processing}
\label{section_backend}

Besides the routine daily updates of the event lists, we perform additional processing to enrich the original data. First, we calculate physical descriptors of the events (coordinates, Temperature and Emission Measure peaks and their times for GOES events) which are in addition to the descriptors already stored in the original lists. Second, we match each event from each primary list with its counterparts in other primary lists, and assign a unique identifier (UniqueID) for each uniquely-matched case. These procedures are described in this section.

\subsection{Determination of Coordinates for the GOES Events}

The GOES flare list reports the events detected from the integrated X-ray light curves and includes coordinates only for some events. However, in most cases, the NOAA active region number where the flare occurred is known and reported in the list, but without its coordinates. To estimate the coordinates of the event based on the active region number, we utilize the Solar Region Summary (SRS) files. Such files are formed every day half-an-hour after midnight and report the current active regions, and their locations at 00:00\,UT. Using these angular coordinates, we compute the position of the active region at the flare start time, assuming the Carrington rotation period $T\approx{}27.3$~days, and taking into account the variations of the solar radius with time as a function of the Earth's position in its orbit.

\subsection{Temperature and Emission Measure for GOES Events}

\begin{figure}[p]
	\centering
	\includegraphics[width=.49\linewidth]{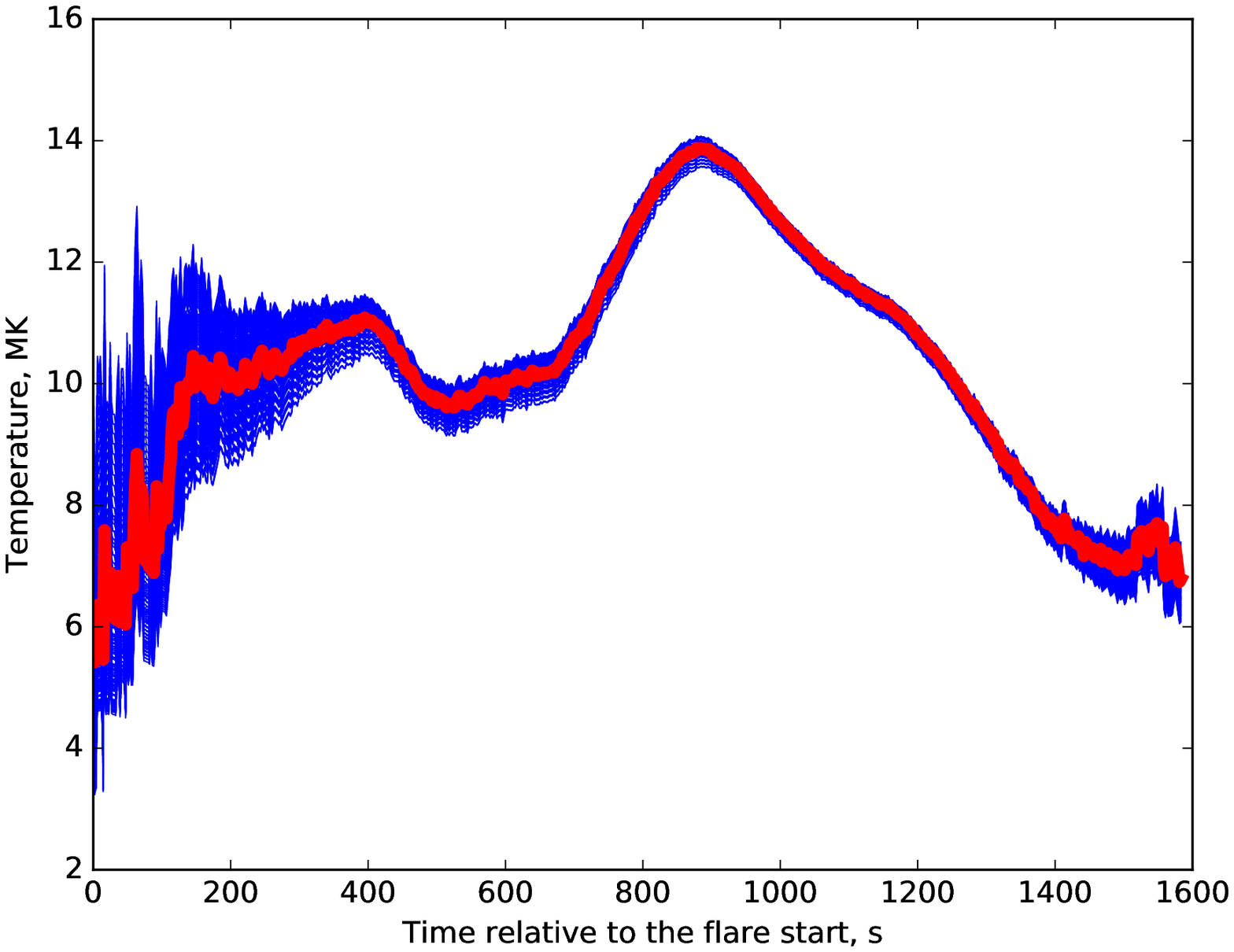}
	\includegraphics[width=.49\linewidth]{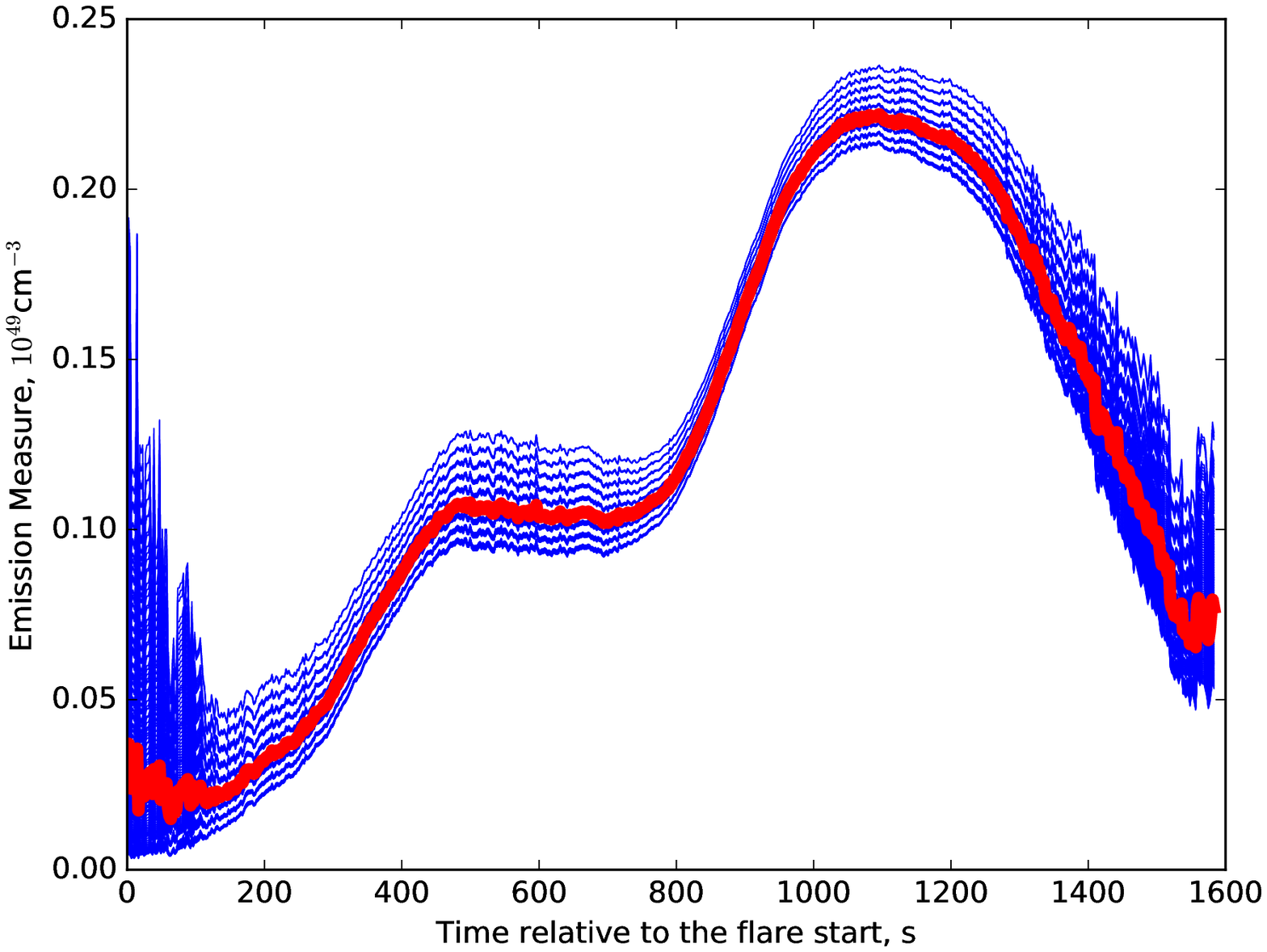}
	\caption{Example of the Temperature (right panel) and Emission Measure (left panel) calculations using the TEBBS algorithm for the SOL2016-02-15T04:02:00 event (C3.9 class flare). The blue curves represent the physically possible Temperature and Emission Measure solutions. The red lightcurves represent the best-estimate solution.}
	\label{tebbs_example}
\end{figure}

Important physical properties derived from the GOES X-ray observations are Temperature (T) and Emission Measure (EM)~\citep{Thomas85,White05}. These parameters can be defined for each moment of time, and provide T and EM profiles for every flare. In our database, we characterize flares by the peak values of these parameters ($T_{max}$ and $EM_{max}$), as well as by the times when these peak values are reached. To remove the background (non-flare) X-ray flux, we use the Temperature and Emission measure Based Background Subtraction (TEBBS) algorithm, initially proposed by~\cite{Bornmann90} and improved by~\cite{Ryan12} based on the assumption that T and EM must grow during the flare impulsive phase. We have implemented in our database the algorithm proposed by~\cite{Ryan12}. The corresponding GitHub repository is available for public access: \url{https://github.com/vsadykov/TEBBS.git}

As mentioned above, the algorithm receives all the physically-possible combinations of the background level, which provide growing T and EM curves after the flare start time. For each of these curves, we calculate the T and EM maximum values during the flare. The range of these values defines the physical interval for $T_{max}$ and $EM_{max}$. To obtain ``the best'' curve representing the T and EM dynamics, we simultaneously minimize the deviation from the $T_{max}$ and $EM_{max}$ median values for all curves, and choose the one corresponding to the minimum mean deviation. For the best estimate curve, we compute $T_{max}$ and $EM_{max}$, and the corresponding time moments, and store them in our database, together with the possible physical intervals of $T_{max}$ and $EM_{max}$. An example of the TEBBS calculations for a C3.9 class flare is presented in Figure~\ref{tebbs_example}.

\subsection{UniqueID Assignment and Relation to the SOL ID}

The three ``primary'' catalogs (GOES, RHESSI and HEK flare lists) of the database are updated on daily basis. For every new flare event, we assign the Unique Identifiers (UniqueIDs) by integrating the information from these three different sources.

Each of the ``primary'' catalogs reports the events with the known start, peak, and end times. Also, the information about the event coordinates is provided or calculated as described above. This information is used to determine if the entries in these catalogs represent the same physical phenomenon, i.e. they happened at the same time in the same place, or they belong to different events.

For assigning the UniqueID, we introduce the following hierarchy order: NOAA GOES, RHESSI, and the HEK flare list. This order means the following: if a flare event is reported by GOES, then it labeled as a GOES event (``gev''). If an event is not in the GOES catalog, but reported by RHESSI, this event is labeled as a RHESSI event (``rhessi''). If an event is not reported by GOES and RHESSI, but recorded as a flare in the HEK database, this is labeled as a HEK event (``hek''). The UniqueID consists of two parts: the name of the primary instrument that observed the flare, and its start time. For example, the GOES event observed at ``yyyy-mm-dd hh:mm'' gets the UniqueID ``gev\_yyyymmdd\_hhmm00'', while the RHESSI or HEK events are labeled ``rhessi\_yyyymmdd\_hhmmss'' and ``hek\_yyyymmdd\_hhmmss'' respectively (we add ``00'' to the GOES event UniqueIDs for compatibility with the UniqueIDs from other catalogs). In the case of events with the same start time but different locations (which was found for some HEK event only), we assign the UniqueID ``hek\_yyyymmdd\_hhmmss\_i'', where ``i'' is an increasing integer, starting from 1 for each such case. The advantage of the event ID assignment procedure we adopted is that this classification can be easily extended for any number of flare-reporting instruments.

The procedure of the UniqueID assignment is the following:
\begin{enumerate}
	\item Query GOES flare list events for their coordinates and active region numbers. Then, sort the events according to their GOES X-ray class in the descending order. For each event (hereafter parental event) assign the UniqueID ``gev\_yyyymmdd\_hhmmss'' according to the event. Then, 
	\subitem a) For each ``gev'' event, find all events in the RHESSI and HEK flare catalogs overlapping in time, from the start to end times, and obtain their coordinates and active regions. This is a list of candidate events corresponding to the parental event.
	\subitem b) For each candidate event compare the coordinates and active regions with the parent event. The events are assigned the same UniqueID as the parent event if they have the same active regions and their location differs by no more than $\delta = 250^{\prime\prime}$. This value was chosen experimentally, and it is approximately equal to the size of a large active region. If one of the compared events (parental or candidate) has coordinate or/and active region information missing, the corresponding condition is assumed to be satisfied.
	\item Repeat the procedure for the events for which the UniqueID is still not assigned, using the RHESSI flare catalog. These events are sorted according to their energy range and their UniqueIDs are assigned in the form ``rhessi\_yyyymmdd\_hhmmss''.
	\item Repeat the procedure for the remaining set of events for which the UniqueID is still not assigned, using the HEK register. The HEK register contains events overlapping in time (for example, reported for different SDO/AIA channels, or from different locations), thus the matching procedure is still needed. For the matched events, assign UniqueIDs ``hek\_yyyymmdd\_hhmmss\_i'', where ``i'' represents a discriminatory index assigned only to those events characterized by the same start time but different locations.
\end{enumerate}

The procedure of the UniqueID assignment for new events is repeated on a daily basis. Nevertheless, a complication may happen if one of the events is reported with a delay of one day or more. If this is the case, the UniqueIDs of the events overlapping with such delayed events are deleted, and the UniqueID assignment starts for all the events with empty UniqueIDs.

The same UniqueID may be defined for several GOES, HEK and RHESSI events. For all such overlapping, we keep the maximum and minimum values for each of the coordinates among all the matched events.

The last thing needed to be mentioned is how our UniqueIDs correspond to other event IDs. The Solar Object Locator (SOL), which in its simplest form contains only the event time, is one of the widely-used identifiers. Because it is not documented whether the event time should correspond to the start, peak or end time, although, for display purposes, we use the flare start time as a default reference time, but we also assign and maintain in the database the correspondence between our UniqueID and these three possible versions of the SOL IDs.

\section{Query Structure and Processing}
\label{section_query}

In this chapter, we discuss the structure of the query engine. This engine is the most important software component for the construction of the Web Application, because it should be efficient, fast, and user-friendly. The current implementation retrieves and displays the final list of events in a convenient form, works fast, and it is constructed in such a way that adding new catalogs does not require changing the code structure.

To perform a query, the user needs to fill the request form in the Web Application site: a web request form available at \url{https://solarflare.njit.edu/webapp.html}. In the request form, the user selects the desired time interval (including the ability to select the whole time range, starting from January 1st, 2002), select the position of the event on the solar disk, apply instrument-specific filters such as event availability of the uniquely-matched events in different catalogs, as well as ranges of various physical parameters, and executes the query by pressing the submit button. Alternatively, the user can load previously saved query result.

\subsection{Primary Catalogs, Filters and Appearance of the Additional Fields}

The ``primary'' flare catalogs (GOES, RHESSI and HEK flare records) are updated on a daily basis and have the flare records detected by their own algorithms. The descriptors of the primary catalogs are displayed independent of the user's selection of filters. For example, if the user is searching for all events listed in the RHESSI flare catalog, the output may have empty GOES or HEK fields. All fields will be populated only if the data availability filters in ``primary'' catalogs are selected.

The on-the-fly generation of additional search fields corresponding to specific event descriptors for the other catalogs depends upon user selection. For example, let us consider the case of a user looking for the events listed in the Hinode catalog. In such cases, the descriptor fields related to the Hinode catalog (number of observed frames, corresponding quicklook link etc.) will appear in the final table, as additional columns. This strategy allows us to make the tables as short and informative as possible based on the user's query. Almost each of the parameter fields may be tuned during the query: the filters allow not only the ability to check the appearance of the event in a certain catalog, but also to select events having particular physical characteristics.

The initially-constructed table (based on the primary catalogs) is the backbone for the query: we simply discard from this table the event records which do not pass additionally selected filters. This allows us to check the selected filters one by one, without pulling them into one large query. This structure has one more advantage: we can add the filters for new-uploaded catalogs/lists without disturbing the working system. Adding of a new event table would just require a new independent block in the query engine.

\subsection{Results Table and Python Routine for Parsing}

The final result of a query  is presented in the form of a web table with moving headers. One can simply drag the table to the right to see various characteristics of the events. For better performance, we currently restrict the number of events appearing in the table to 1000. However, the full lists of events are available for downloading in the output file, which can be saved locally and reloaded anytime. We also added the possibility to sort the output table according to the flare characteristics. The sorting procedure represents another query to the server and includes all events, even if the number of events exceeds 1000. The user has the option to download the output table. In order to simplify the processing of the output file, we created a Python parser which reads the table and creates the structure corresponding to the events.

\subsection{Detailed Visualization of a Selected Event}

The main purpose of the created database is to integrate the entries from different catalogs, and to present a complete list of events satisfying a set of conditions specified by the user. However, from a practical perspective, it is very important to have a brief look at the event data, and decide whether a particular event is interesting for a case study or not. For this purpose, we created the ability to look at the event light curves derived from different instruments. To proceed into the event page, one needs to select the event of interest from the summary table retrieved in the previous stage, and click the ``Plot Data'' button.

The main elements of this event page are the two dynamic graphs reflecting the behavior of several event light curves: X-ray fluxes, Temperature and Emission Measure calculated for the GOES data using the TEBBS algorithm, light curves from the SDO/EVE/ESP instrument, and the Nobeyama Radio Polarimeter fluxes. The user can select which plot to display, and scale it accordingly. For visualization, we are currently using the Google Charts tool.

The interactive web interface also allows the user to download all the displayed light curves. The downloaded file contains the GOES data with 2s resolution, and the 10s averaged Nobeyama and SDO/EVE/ESP data. Besides the graphs, we also provide the user with an image of the flare generated from SDO/AIA 1600\,{\AA} data, and a detailed description of all overlapping events from the primary GOES, RHESSI and HEK lists corresponding to the same UniqueID, as well as from the secondary event sources. Besides the usual flare descriptors, the HEK database contains links to the flare quicklook images: we keep these links in our event summary page, which also can be downloaded.

In addition, the users are provided with the option of a similarity search mechanism (currently beta version) that allows automatic selection of similar events from the initial query table, based on some predefined and user-defined characteristics. Each such event characteristic is normalized, and, if the associated fields are absent from the table for some of the events, they are replaced by the median values of the corresponding characteristic. The nearest neighbors of the selected event are determined based on the selected (predefined) characteristics using the Euclidean distance.

\subsection{Example of a Query}

To demonstrate the capabilities of our Multi-Instrument Database of Solar Flares, we provide here an example of a multi-instrument query. Suppose that the user wants to study the chromospheric evaporation processes occurred during strong solar flares ($\geq$M1.0) in 2015 using the spatially-resolved high-cadence multiline spectroscopic observations performed by the IRIS satellite, and simultaneously analyze the energy released by precipitating accelerated electrons using RHESSI observations. Selecting the corresponding GOES filter: GOES class $\geq$M1.0; the IRIS filter: expansion of the field-of-view by 100$^{\prime{}\prime{}}$, $\geq$4 slit positions, $\geq$5$^{\prime{}\prime{}}$ covered perpendicular to the slits with cadence of $\leq$60s; the availability of RHESSI observations, and selecting the non-limb events (located not farther than 750$^{\prime{}\prime{}}$ from the disc center to avoid strong projection effects), the query would return six records. Some descriptors of these flares are presented in Table~\ref{table_query}. After such a query, the user can check the events manually: see the lightcurves for each event, proceed to the IRIS quick look images and movies, and check how well the events were covered by the IRIS slit positions, etc.

\section{Conclusion and Future Plans}
\label{section_conclusion}

We have created an Interactive Multi-Instrument Database of the Solar Flares available to the community at \url{ https://solarflare.njit.edu/}. This database integrates a set of available solar flare lists and data in a convenient way, and includes the following main features:

\begin{itemize} 	
	\item The integration of the flare events from different flare catalogs (GOES, RHESSI, HEK, Hinode, Fermi GBM, Konus-Wind, OVSA flare catalogs). The match of events from GOES, RHESSI and HEK primary flare lists, and assignment of Unique Identifiers (UniqueIDs) for flares. The queries provide "one flare~--- one result." After the UniqueID assignment, the flare reports are integrated with secondary flare catalogs (Hinode, Fermi GBM, Konus-Wind, OVSA) and flare-related events (Filament Eruption catalog, CACTus CME catalog), depending on the user's query.
	\item The search of the flare events based on their physical descriptors (both stored in the catalogs and calculated by our efforts) and availability of observations (currently IRIS and Nobeyama observational filters are available). The search allows the users to select the events of interest based on the specified filters, get the integrated properties of the events in one table, download the results of the query, and visualize the processed light curves for each event.
	\item The detailed look at the data (GOES, ESP/EVE and Nobeyama light curves, and temperature and emission measure derived from GOES data) for a particular event, and to its summary containing quicklook links stored in the primary catalogs, allows the user to form an initial opinion about the selected event, and to decide whether the event would be interesting for a case study.
\end{itemize}

The integrated catalog results generated by our database provide a tool to assist researches who study solar flares using large data archives. Firstly, the tool we have created allows the user to search for events having the parameters of interest for various statistical studies, handling all the catalog-creation tasks, or at least providing a catalog to start from. Secondly, it provides a summary for each event, allowing the researchers to understand if the particular event satisfies the criteria for particular case studies. Our web application allows a platform- and software-independent access to the data.

As far as we know, there are almost no examples of such kinds of query engines for solar flares. In this case, our database really provides a unique overlook of the flare data. Currently, there are many filters, catalogs and data processing modules already implemented in our database. However, the design allows further addition of the instrumental logs and sources without distortion of the current schema. Further expansion of the sources is definitely in our plans. We also plan to increase the flexibility of the project by developing a true Web API which will allow the user to receive the flare lists, apply different integration schema, and contribute to the database by adding his own records and data. We also plan to integrate the VSO API and generate data links for the stored flare events.

\acknowledgments
We thank the anonymous referee for the extremely helpful and detailed review of the paper. We thank Rishabh Gupta and Nalinaksh Gaur for their help in developing the web interface. We thank teams of the GOES, RHESSI, SDO, IRIS, Fermi and Hinode space missions, and also OVSA and Nobeyama Radio observatories for the availability of the high-quality scientific data. We also thank the teams managing the currently used catalogs (GOES, RHESSI, Hinode, Fermi GBM, Konus-WIND, OVSA flare catalogs, Filament eruption catalog, CACTus CME catalog and Heliophysics Event Knowledgebase) for the possibility to work with their data. The research was partially supported by the NASA Grants NNX15AN48G, NNX14AB68G and NNX16AP05H, and by the NJIT Faculty Seed Grant (FSG).

\bibliographystyle{aasjournal}

\bibliography{Solarflare.njit.edu}

\clearpage

\begin{figure}[p]
	\centering
	\includegraphics[width=0.8\linewidth]{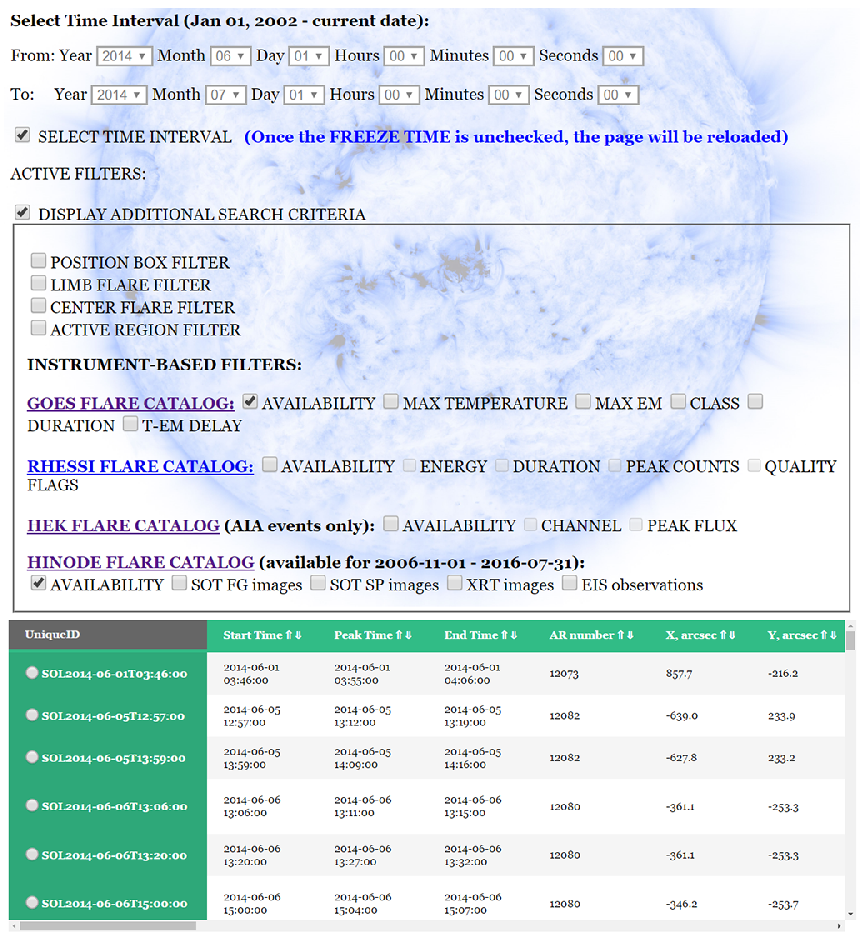}
	\caption{Example of the query for the flare events detected in both the GOES flare list and Hinode flare catalog of June, 2014.}
	\label{query_example}
\end{figure}

\clearpage
\begin{table}
	{\footnotesize
	\caption{Event catalogs currently implemented in the Interactive Multi-Instrument Database of Solar Flares (https://solarflare.njit.edu/).}
	\label{table_datalinks}
	\begin{tabular}{ccc}
		\hline
		Source Name	&	Dates presented	&	Source web link \\
		\hline
		\multicolumn{3}{c}{Primary flare lists} \\
		GOES flare list	&	Jan, 2002~--- current time	&	\url{ftp://ftp.swpc.noaa.gov/pub/warehouse/} \\
		RHESSI flare list	&	Feb, 2002~--- current time	&	\url{http://hesperia.gsfc.nasa.gov/hessidata/dbase/} \\
		HEK flare list	&	Feb, 2010~--- current time	&	\url{https://www.lmsal.com/isolsearch} \\
		\hline
		\multicolumn{3}{c}{Secondary event catalogs} \\
		IRIS observing logs	&	Jul, 2013~--- current time	&	\url{http://iris.lmsal.com/search/} \\
		Hinode flare catalog	&	Nov, 2006~--- July, 2016	&	\url{http://st4a.stelab.nagoya-u.ac.jp/hinode_flare/} \\
		Fermi GBM flare catalog	&	Nov, 2008~--- current time	&	\url{https://hesperia.gsfc.nasa.gov/fermi/gbm/qlook/} \\
		Nobeyama coverage check	&	Jan, 2010~--- current time	&	\url{ftp://solar-pub.nao.ac.jp/pub/nsro/norp/xdr/} \\
		OVSA flare catalog	&	Jan, 2002~--- Dec, 2003	&	\url{http://www.ovsa.njit.edu/data/} \\
		CACTus CME catalog	&	Jan, 2002~--- current time	&	\url{http://sidc.oma.be/cactus/} \\
		Filament eruption catalog	&	Apr, 2010~--- Oct, 2014	&	\url{http://aia.cfa.harvard.edu/filament/} \\
		Konus-Wind flare catalog	&	Jan, 2002~--- Jul, 2016	&	\url{http://www.ioffe.ru/LEA/Solar/index.html} \\
		\hline
	\end{tabular}
	}
\end{table}

\clearpage
\begin{table}
	\caption{Results of the sample query (see text for details).}
	\label{table_query}
	\begin{tabular}{cccc}
		\hline
		SOL ID	&	Flare Class	&	RHESSI highest energy	&	IRIS raster mode  \\
			&		&		&	and number of slit positions \\
		\hline
		SOL2015-03-10T23:46:00	&	M2.9	&	12-25\,keV	&	coarse, 4-step \\
		SOL2015-03-11T16:11:00	&	X2.1	&	25-50\,keV	&	coarse, 4-step \\
		SOL2015-03-11T18:37:00	&	M1.0	&	25-50\,keV	&	coarse, 4-step \\
		SOL2015-06-22T17:39:00	&	M6.5	&	100-300\,keV	&	sparse, 16-step \\
		SOL2015-08-27T04:48:00	&	M2.9	&	12-25\,keV	&	coarse, 8-step \\
		SOL2015-11-04T13:31:00	&	M3.7	&	50-100\,keV	&	coarse, 16-step \\
		\hline
	\end{tabular}
\end{table}

\end{document}